\documentclass[journal]{IEEEtran}
\usepackage{mathrsfs,amsfonts,amsmath,amssymb,graphicx,subfigure,algorithm,algorithmic,txfonts}

\newcommand{\RR}{{\mathbb R}}
\newcommand{\PP}{{\mathcal{P}}}
\newcommand{\DD}{{\mathcal{D}}}
\newcommand{\OO}{{\mathcal{O}}}

\newcommand{\XX}{{\mathbf{X}}}
\newcommand{\CC}{{\mathbf{C}}}
\newcommand{\oo}{{\mathbf{1}}}
\newcommand{\AAA}{{\mathbf{A}}}
\newcommand{\YY}{{\mathbf{Y}}}
\newcommand{\no}{\nonumber}

\newcommand{\be}{\begin{eqnarray}}
\newcommand{\ben}{\begin{eqnarray*}}
\newcommand{\en}{\end{eqnarray}}
\newcommand{\enn}{\end{eqnarray*}}
\newcommand{\bef}{\begin{figure}\begin{center}}
\newcommand{\befd}{\begin{figure*}\begin{center}}
\newcommand{\enf}{\end{center}\end{figure}}
\newcommand{\enfd}{\end{center}\end{figure*}}

\newtheorem{theorem}{\bf{Theorem}}[section]

\begin{document}
\title{A Weighted Common Subgraph Matching Algorithm}
\author{Xu Yang, \emph{Student Member}, Hong Qiao, \emph{Senior Member}, and Zhi-Yong Liu, \emph{Member}, IEEE
\thanks{Xu Yang, Hong Qiao and Zhi-Yong Liu are with the State Key Laboratory of Management and Control for Complex Systems,
Institute of Automation, Chinese Academy of Sciences. Corresponding author: Zhi-Yong Liu (zhiyong.liu@ia.ac.cn)}}
\IEEEtitleabstractindextext{
\begin{abstract}
We propose a weighted common subgraph (WCS) matching algorithm to find the most similar subgraphs in two labeled weighted graphs. WCS matching, as a natural generalization of the equal-sized graph matching or subgraph matching, finds wide applications in many computer vision and machine learning tasks. In this paper, the WCS matching is first formulated as a combinatorial optimization problem over the set of partial permutation matrices. Then it is approximately solved by a recently proposed combinatorial optimization framework - Graduated NonConvexity and Concavity Procedure (GNCCP). Experimental comparisons on both synthetic graphs and real world images validate its robustness against noise level, problem size, outlier number, and edge density.
\end{abstract}
}
\maketitle
\IEEEdisplaynontitleabstractindextext
\section{Introduction}
Graph matching aims to find the optimal correspondence between vertices of two graphs. It is a fundamental problem in theoretical computer sciences, and also plays a key role in many computer vision and machine learning tasks, such as object recognition and feature correspondence.

Bipartite graph matching can be effectively and efficiently solved by the Hungarian algorithm \cite{Hungarian} or linear programming methods \cite{assignment}. When further considering pairwise constraints, the matching problem becomes NP-hard. Approximate methods which make certain relaxations to the original problem are necessary for efficiency reasons \cite{thirty}.

In the last ten years, significant progresses have been achieved on the approximate methods. For instance, the computational complexity has been decreased to as low as $\OO(N^3)$ - the complexity of matrix multiplication. On the other hand, the accuracy, taking the benchmark dataset `House sequence'\footnote{Available at http://vasc.ri.cmu.edu//idb/html/motion/house/index.html} for example, has been increased from about 60\% \cite{counter_example} to nearly 100\% \cite{fgm, path}. The progresses are mainly due to the introduction of proper graph similarity criterions and optimization techniques. Typical algorithms in literature include graduated assignment \cite{ga}, spectral technique \cite{spectral}, path following \cite{path}, probabilistic graph matching \cite{pgm}.

In this paper we consider the matching problem involving outliers. Most existing methods treat it as a part-in-whole problem \cite{sub_part}, commonly known as \emph{subgraph matching} \cite{JMLR_liu} which recognizes the smaller graph as a part of the bigger one. Moreover, some recently proposed effective graph matching algorithms \cite{path, epath} are only applicable on equal-sized graphs. However, in realistic computer vision applications, there may exist outliers in both images because of image background, object occlusion or geometric transformations. Thus it is reasonable to formulate the matching problem as finding the most similar subgraphs within two graphs abstracted from the images. Furthermore, to obtain a robust similarity measure, the number of matched vertices should sometimes be specified and kept lower than the estimated number of inliers representing the objects \cite{robust_point}. Then the problem can be defined as finding the most similar subgraphs with a specified size in two labeled weighted graphs in some optimal way. We denote the problem by weighted common subgraph (WCS) matching, which can be taken as a generalization of the equal-sized graph matching and subgraph matching problems.

Another similar term in the literature is the maximum common subgraph (MCS) problem, or known as maximum common subgraph isomorphism \cite{MCS}. Given two graphs, MCS aims to find the largest subgraph in one graph isomorphic to an unknown subgraph in the other graph. MCS has a long tradition in structural data processing, such as cheminformatics. MCS and WCS are different mainly from the following two aspects. First, MCS requires the two common subgraphs to be strictly isomorphic to each other, even on weighted graphs, while WCS tolerates some disparities between them. The latter one is more reasonable in most computer vision tasks. Second, MCS searches for the largest common subgraphs while WCS for the common subgraphs of a specified size.

In the literature, there exist some algorithms \cite{spectral, pgm} applicable to the WCS matching, by typically first matching all the vertices, and then finding a specified number of best assignments by ranking techniques. Unfortunately, such a two-step idea is not completely consistent with the original WCS problem. That is, even both the two steps are optimally solved, the obtained subgraphs may not be the optimal pair.

Different from the above methods, in this paper we propose a novel WCS algorithm which unifies the two steps and targets directly at the specified number of best assignments. Specifically, the contributions of this paper are two-fold. The first one is to formulate the WCS problem as a combinatorial problem over the partial permutation matrices. The second one is to develop a GNCCP \cite{GNCCP, NC_partial} based optimization algorithm, for which we propose two relaxations of the objective function to make the calculation tractable.

The remaining of the paper is organized as follows. The WCS matching algorithm is proposed in Section \ref{sec:method}. Then it is experimentally evaluated in Section \ref{sec:experiment} on both synthetic graphs and real world images. Finally the concluding remarks and future extensions are discussed in Section \ref{sec:conclusion}.

\section{Method}\label{sec:method}
In this section, we first formulate the WCS problem as a combinatorial optimization problem, and then approximately solve it by the GNCCP. Finally we give two relaxations to make the algorithm implementable.
\subsection{Formulation}
A graph $G=(V,E)$ of size $M$ is defined by a finite vertex set $V = \{1,2,\cdots,M\}$ and an edge set $E \subseteq V\times V$. The labeled weighted graph is further defined by assigning a real number vector $l^G_i$ as a label to vertex $i$, and assigning a nonnegative real number $w^G_{ij}$ as a weight to edge $ij$ in $G$. Taking feature correspondence for example, by treating feature points as vertices, some local descriptor, e.g. SIFT descriptor, can be used as the vertex label, and the distance between two feature points as the edge weight. The weighted adjacency matrix $\AAA_G\in\RR^{M\times M}$ is commonly used to record adjacency and weights of edges. Hereafter by terms \emph{graph} and \emph{adjacency matrix}, we mean the labeled weighted graph and weighted adjacency matrix respectively.

Then, the WCS matching problem is formally described as follows.\\
\emph{WCS(G,H)}\\
\textbf{Input:} graph $G$ of size $M$, graph $H$ of size $N$, and an integer $L$. Assume $L\leq M\leq N$.\\
\textbf{Question:} Which subgraph of size $L$ in $G$ is most similar to an unknown subgraph of size $L$ in $H$ under certain optimal criterions? And what is the optimal correspondence between vertices of the two subgraphs?

\begin{figure}[ht]
\begin{center}
%\vskip 0.2in
\includegraphics[scale=0.5]{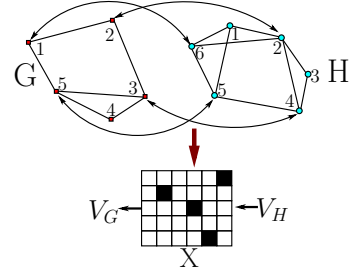}
\caption{Matching $L=4$ vertices between the graphs $G$ and $H$ with size $M=5$ and $N=6$. The black box and white box in $\XX$ mean $1$ and $0$ respectively.}\label{Fig:formulation}
\end{center}
\vskip -0.2in
\end{figure}

\begin{figure}[ht]
\begin{center}
\includegraphics[width = .99\linewidth, bb = 126 418 524 663]{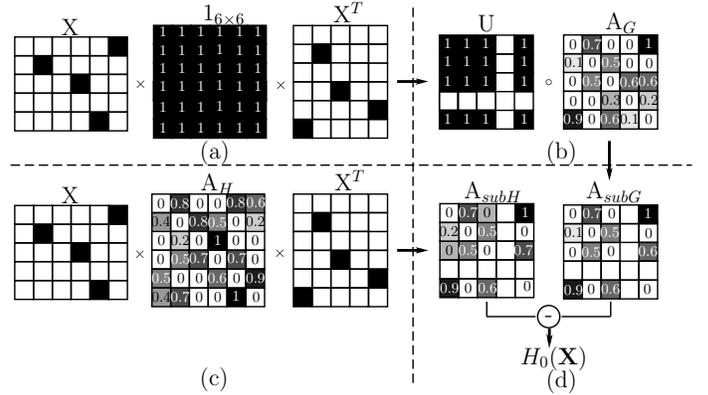}
\caption{The graphic description for the objective. Here $M = 5$, $N = 6$ and $L = 4$. $\AAA_{subG}$ and $\AAA_{subH}$ denote the adjacency matrices for the subgraphs from $G$ and $H$ respectively. (a) $\mathbf{U}=\XX\oo_{N\times N}\XX^T$; (b) $\AAA_{subG} = \mathbf{U}\circ \AAA_G$; (c) $\AAA_{subH} = \XX \AAA_H\XX^T$; (d) $F(\XX)= \|\AAA_{subG} - \AAA_{subH}\|^2_{F}$}\label{Fig:formulation2}
\end{center}
\vskip -0.1in
\end{figure}

WCS can be then formulated as the following combinatorial programming problem:
\be \label{eqn:obj}
&&\hspace{-2cm}\min_\XX F(\XX),\\
&&\hspace{-2cm}\text{s.t.~}\XX\in \PP, \PP:=\left\{\mathbf{X}|\sum_{i=1}^M{\mathbf{X}_{ij}} \leq 1, \sum_{j=1}^N{\mathbf{X}_{ij}} \leq 1,\right. \no\\ &&\hspace{-1.5cm}\left.\sum_{i=1}^M\sum_{j=1}^N{\mathbf{X}_{ij}} = L, \mathbf{X}_{ij}\in\{0, 1\}\right\}, L\leq M\leq N,\no
\en
where $\PP$ is the set of partial permutation matrices illustrated by Fig. \ref{Fig:formulation}, and the objective function $F(\XX)$ is given by
\be
\label{obj0}
F(\XX) &=& \alpha\|\mathbf{U} \circ \AAA_G - \XX \AAA_H\XX^T\|_F^2 + (1-\alpha)\text{tr}(\CC^T\XX) \no \\
&=&\alpha H_0(\XX) + (1-\alpha)\text{tr}(\CC^T\XX).
\en
The Frobenius matrix norm denoted by $\|\cdot\|_F$ is defined as $\|\AAA\|_{F} = \sqrt{\sum_i\sum_j\AAA_{ij}^2}=\sqrt{\text{tr}(\AAA^T\AAA)}$, where $\text{tr}(\cdot)$ denotes the matrix trace. The adjacency matrices $\AAA_G, \AAA_H\in\RR^{N\times N}$ are respectively associated with the graphs $G$ and $H$.  The Hadamard product (entry-wise product) of two matrices, denoted by $\circ$, is defined as $(\AAA\circ \mathbf{B})_{ij} = \AAA_{ij}\mathbf{B}_{ij}$ assuming conformability. The matrix $\mathbf{U}=\XX\oo_{N\times N}\XX^T$ is to `pick out' the vertices with corresponding relations in $G$, where every entry in $\oo_{N\times N}$ is `1', as illustrated by Fig. \ref{Fig:formulation2}. The higher order term in (\ref{obj0}) is denoted by $H_0(\XX)$ for further derivation convenience. In the unary term $\text{tr}(\CC^T\XX)$, $\CC$ is a cost matrix where $\CC_{ij}$ measures the dissimilarity between the labels $l^G_i$ and $l^H_j$. The parameter $\alpha$ is used to balance the two terms.

When $L=M\leq N$, the WCS matching degenerates to the part-in-whole problem \cite{GNCCP}. When $L=M=N$, it degenerates to the equal-sized matching problem \cite{path, epath}.

By defining the problem over $\PP$, we actually introduce the one-to-one constraints on the WCS matching, which is a common assumption in graph matching \cite{fgm}. Particularly, if $\XX_{ij} = 1$, vertex $i$ in $G$ is assigned to $j$ in $H$. If $\sum_i^M \XX_{ij}= 0$, there are no corresponding vertices in $G$ for $j$ in $H$. It is similar when $\sum_j^N \XX_{ij}= 0$, as illustrated by Fig. \ref{Fig:formulation}.

The formulation (\ref{eqn:obj}) directly targets at the WCS matching problem, without resorting to the conventional two-step schema \cite{spectral, pgm}. However, its objective function becomes more complicated to handle, as discussed later in Section \ref{subsection:obj}.

\subsection{GNCCP Based Algorithm}
The combinatorial optimization problem (\ref{eqn:obj}) is NP-hard with a factorial complexity, which calls for some approximations in realistic applications. Below we propose an approximate algorithm based on the GNCCP \cite{GNCCP, NC_partial}, a relaxation technique.

The GNCCP has its root in the convex-concave relaxation procedures (CCRP) \cite{path, epath, JMLR_liu, Liu_IJCV}. Combining both convex and concave relaxations, CCRP achieved a superior performance on the equal-sized graph matching. The GNCCP realizes exactly a type of CCRP, but in a much simpler way. It does not need to construct the convex or concave relaxations explicitly. This is particularly important for the WCS matching because both the convex and concave relaxations are difficult to construct.

To utilize the GNCCP to solve (\ref{eqn:obj}), firstly we need to get the convex hull $\DD$ of $\PP$ as follows,
\begin{theorem}\label{theorem:convex}
The convex hull of the set of partial matrices $\PP$ is $\DD$, where
\ben
\DD:=\left\{\mathbf{X}|\sum_{i=1}^M{\mathbf{X}_{ij}} \leq 1, \sum_{j=1}^N{\mathbf{X}_{ij}} \leq 1, \sum_{i=1}^M\sum_{j=1}^N{\mathbf{X}_{ij}} = L, \mathbf{X}_{ij}\geq 0\right\}.
\enn
\end{theorem}
\emph{Proof:} See Appendix A in the supplementary materials.

Note that $\DD$ can be regarded as a generalization of the set of doubly stochastic matrices \cite{convex_analysis}, the convex hull of the set of permutation matrices. Then the GNCCP takes the following form:
\be\label{GNCCP}
&&\hspace{-1.3cm}J_{\zeta}(\XX)=\begin{cases}
(1-\zeta)F(\XX)+\zeta\text{tr}(\XX^T\XX), \text{ if } 1\geq \zeta\geq 0,\\
(1+\zeta)F(\XX)+\zeta\text{tr}(\XX^T\XX), \text{ if } 0> \zeta\geq -1,
\end{cases}\\
&&\hspace{-1.3cm}\XX\in\DD.\no
\en

In implementation, $\zeta$ decreases gradually from $1$ to $0$ (Graduated NonConvexity) and finally to $-1$ (Graduated Concavity). During the process, GNCCP implicitly realizes the transition from the convex relaxation to the concave relaxation. When reaching the concave relaxation, the continuous minimum point is finally pushed into $\PP$, because $\PP$ is the extreme point set of $\DD$, as indicated by Theorem \ref{theorem:convex}.

For a specific $\zeta$, $J_{\zeta}(P)$ is optimized by the Frank-Wolfe (FW) algorithm \cite{FW}, which iteratively updates $\XX$ by $\XX \leftarrow \XX + \lambda\text{d}$ until converged. The initial $\XX$ is the solution of $J_\zeta(\XX)$ obtained at the previous $\zeta$. And the optimal search direction $\text{d} = \YY-\XX$ is given by solving the following linear programming problem:
\be\label{linear_programming}
&&\hspace{-4.5cm} \YY = \arg \max \text{tr}(-\nabla J_{\zeta}(\XX)^T\YY), \\
&&\hspace{-4.5cm}\text{s.t.~}\YY \in \mathcal{D},\no
\en
which can be solved by, for example, the interior point method \cite{convex_optimization}. The gradient $\nabla J_{\zeta}(\XX)$ in (\ref{linear_programming}) takes the following form:
\be\label{gradient}
&&\hspace{-0.7cm}\nabla J_{\zeta}(\XX)=\begin{cases}
(1-\zeta)\nabla F(\XX)+2\zeta \XX, \text{ if } 1\geq \zeta\geq 0,\\
(1+\zeta)\nabla F(\XX)+2\zeta \XX, \text{ if } 0> \zeta\geq -1,
\end{cases}\\
&&\hspace{-0.7cm}\XX\in\DD.\no
\en
where
\be
\hspace{-2.7cm}\nabla F(\XX) = \nabla H_0(\XX) + (1-\alpha)\CC.
\en
The optimal step size $\lambda$ is given by
\be
&&\hspace{-4.3cm}\lambda = \arg \min J_{\zeta}(\XX + \lambda(\YY - \XX)), \\
&&\hspace{-4.3cm}\text{s.t.~}0\leq\lambda\leq 1 \no,
\en
which can be solved by inexact line search, e.g. backtracking algorithm \cite{convex_optimization}.

Finally, the GNCCP based WCS matching algorithm is summarized as follows,
\begin{description}
  \item[\textbf{Input:} Two graphs $G$ and $H$]
  \item[\textbf{Initialization:} $\XX \leftarrow \oo_{M\times N}\frac{L}{M\times N}, \zeta\leftarrow 1$]
  \item[\textbf{GNCCP:}]~
    \begin{description}
      \item[\textbf{Repeat}] ~\\
      FW process\\
      $\zeta = \zeta - d\zeta$
      \item[\textbf{Until}] $\zeta < -1 \vee \XX\in\PP$
    \end{description}
  \item[\textbf{Output:} The matching result $\XX$]
\end{description}

\subsection{Implementation Details}
\label{subsection:obj}

In implementation, it is difficult to directly calculate $\nabla H_0(\XX)$ which involves a Hadamard product. Instead, below we propose two types of relaxations of $H_0(\XX)$ to make it calculable. Similar relaxation techniques were widely used in graph matching \cite{path, epath}.

The first relaxation $H_1(\XX)$ is given as follows:
\be \label{obj1}
&&\hspace{-2.2cm}H_1(\XX) = \|\mathbf{U} \circ \AAA_G - \XX \AAA_H\XX^T\|_F^2 = \text{tr}((\AAA_G \circ \AAA_G)U^T) \no\\
&&\hspace{-2.2cm}- 2\text{tr}(\AAA_G\XX \AAA_H^T\XX^T) + \text{tr}(\XX \AAA_H\XX^T\XX \AAA_H^T\XX^T),
\en
where we take advantage of
\begin{subequations}
\label{trans_2}
\begin{equation}
\mathbf{U} \circ \mathbf{U} = \mathbf{U},
\end{equation}
\begin{equation}
\mathbf{U} \circ (\XX \AAA_H \XX^T) = \XX \AAA_H \XX^T.
\end{equation}
\end{subequations}
See Appendix B in the supplementary materials for the derivation details of (\ref{obj1}). Its gradient is then figured out as follows:
\be
&&\hspace{-1.7cm}\nabla H_1(\XX) = (\AAA_G^T \circ \AAA_G^T + \AAA_G \circ \AAA_G)\XX \mathbf{1}_{N\times N} - 2(\AAA_G^T\XX \AAA_H \no\\
&&\hspace{-1.7cm} + \AAA_G\XX \AAA_H^T) + 2(\XX \AAA_H\XX^T\XX \AAA_H^T + \XX \AAA_H^T \XX^T\XX \AAA_H).
\en
The second relaxation $H_2(\XX)$ is derived as follows:
\be\label{obj2}
&&\hspace{-0.8cm}H_2(\XX) = \|\mathbf{U} \circ \AAA_G- \XX \AAA_H\XX^T\|_F^2 = \|(\XX\XX^T)\AAA_G(\XX\XX^T)- \XX \AAA_H\XX^T\|_F^2\no\\
&&\hspace{-0.8cm}=\text{tr}(\XX\XX^T\AAA_G^T\XX\XX^T\AAA_G\XX\XX^T)-2\text{tr}(\XX\XX^T\AAA_G^T\XX \AAA_H\XX^T)\no\\
&&\hspace{-0.8cm}+\text{tr}(\XX \AAA_H^T\XX^T\XX \AAA_H\XX^T)=T_1(\XX) - 2T_2(\XX) + T_3(\XX),
\en
where we take advantage of
\begin{subequations}
\label{trans_1}
\begin{equation}
U\circ A = \XX\XX^TA\XX\XX^T
\end{equation}
\begin{equation}
\XX\XX^T\XX\XX^T = \XX\XX^T
\end{equation}
\begin{equation}
\XX\XX^T\XX = \XX
\end{equation}
\begin{equation}
\XX^T\XX\XX^T = \XX^T.
\end{equation}
\end{subequations}
Then the gradient is given as follows:
\be
\nabla H_2(\XX)=\nabla T_1(\XX) - 2\nabla T_2(\XX) + \nabla T_3(\XX),
\en
where
\ben
\nabla T_1(\XX)=2(\XX\XX^T\AAA_G^T\XX\XX^T\AAA_G\XX + \AAA_G^T\XX\XX^T\AAA_G\XX\XX^T\XX \\
+ \AAA_G\XX\XX^T\XX\XX^T\AAA_G^T\XX),
\enn
\ben
\nabla T_2(\XX)=\XX \AAA_H^T\XX^T\AAA_G\XX + \AAA_G^T\XX \AAA_H\XX^T\XX + \AAA_G\XX\XX^T\XX \AAA_H^T\\
 + \XX\XX^T\AAA_G^T\XX \AAA_H,
\enn
\ben
\nabla T_3(\XX)=2(\XX \AAA_H^T\XX^T\XX \AAA_H + \XX \AAA_H\XX^T\XX \AAA_H^T).
\enn

Consequently, by replacing $H_0(\XX)$ with $H_1(\XX)$ or $H_2(\XX)$, the GNCCP can be implemented to solve the WCS matching problem. It is noted that both $H_1(\XX)$ and $H_2(\XX)$ are relaxations of $H_0(\XX)$ because $H_0(\XX) = H_1(\XX)=H_2(\XX),~\forall \XX\in\PP$. However, the equivalence becomes in general unsatisfied when $\XX\in\DD\setminus \PP$.

It is difficult to evaluate the two relaxations theoretically by such as the error bound because neither of them is convex relaxation. However, as revealed by the experimental comparisons in section \ref{sec:experiment}, $H_1(\XX)$ outperforms $H_2(\XX)$ in most of the results. The advantage of  $H_1(\XX)$ is probably due to the fact that its order is the same as $H_0(\XX)$, and is lower than that of $H_2(\XX)$. Furthermore, $H_1(\XX)$ is computationally more efficient than $H_2(\XX)$.

Last but not least, when degenerating to the part-in-whole problem, i.e., $L=M$, the GNCCP can be directly implemented, with $H_0(\XX)$ and its gradient $\nabla H_0(\XX)$ given as follows \cite{GNCCP}:
\be\label{obj3} H'_0(\XX) = \|\AAA_G - \XX \AAA_H\XX^T\|_F^2 \en
\be\label{dobj3} \nabla H'_0(\XX) = 2\XX(\AAA_H^T\XX^T\XX \AAA_H + \AAA_H\XX^T\XX \AAA_H^T) \no\\
- 2(\AAA_G\XX \AAA_H^T + \AAA_G^T\XX \AAA_H). \en

\subsection{Storage and Computational Complexity} \label{fast_method}
The proposed method formulates the graph matching problem based on the adjacency matrix. Compared with the \emph{affinity matrix}\footnote{The affinity matrix can be seen as the adjacency matrix for the associate graph of $G$ and $H$, whose size is $MN\times MN$.} based algorithms \cite{ga, spectral, balanced, pgm, fgm}, one most important advantage of the adjacency matrix based methods is storage saving. Without considering the sparsity, the storage complexity of the affinity matrix based methods is $\OO(M^2N^2)$, while that of adjacency matrix based methods, including the proposed method, is as low as $\OO(N^2)$.

The computational complexity of the proposed method is mainly determined by the linear programming problem (\ref{linear_programming}), which can be solved in polynomial time. When $L=M\leq N$, (\ref{linear_programming}) can be solved by, for example, the rectangle Hungarian algorithm \cite{PHungarian} with an $\OO(M^2N)$ computational complexity. It is smaller than the matrix multiplication complexity $\OO(MN^2)$, so the overall complexity is $\OO(MN^2)$. When $L<M\leq N$  which makes the Hungarian algorithm or some efficient linear assignment algorithms \cite{assignment} inapplicable, (\ref{linear_programming}) usually resorts to some general linear programming algorithms involving an $\OO(N^6)$ complexity, such as the interior point method.

To make the algorithm more efficient, a fast method is presented to approximately solve (\ref{linear_programming}), which maintains the complexity of proposed method as $\OO(MN^2)$. The fast method first finds a solution $\YY_1$ for (\ref{linear_programming}) with $M$ `1's by the rectangle Hungarian algorithm \cite{PHungarian}. Then from $\YY_1$ it removes $M-L$ `1's which correspond to the $M-L$ smallest values in all the $M$ corresponding values in $-\nabla J_\zeta(\XX)$. Finally an approximate solution $\YY$ with $L$ `1's is obtained. The cost for the computational efficiency is some loss of matching accuracy, as to be experimentally demonstrated in the next section.

\section{Experimental Results}\label{sec:experiment}
We apply the proposed algorithm on synthetic graphs as well as real world images \footnote{More experimental results are given in the supplementary materials (Sup\_Mat\_II\_Add\_Exp.pdf), including an experiment on handwritten Chinese character recognition and typical results on \emph{Motorbike} and \emph{Pisa} images.}, to evaluate its performance against noise level, problem size, outlier number, and edge density. The experiments are conducted on both WCS and part-in-whole (PIW) problems. The methods included for comparison are \emph{Spectral technique} (SM) \cite{spectral}, \emph{Graduated assignment} (GA) \cite{ga}, \emph{Probabilistic graph matching} (PGM) \cite{pgm}, \emph{(Extended) Path following method} (EPF) \cite{path, epath}. The proposed algorithm with two relaxations are denoted by RLX1 and RLX2 respectively. When the fast method described in \ref{fast_method} is used to solve the linear programming (\ref{linear_programming}), the two algorithms are then respectively denoted by RLX1F and RLX2F. When used on the PIW problem, the proposed algorithm is denoted by OUR.

%becomes exactly the one proposed in \cite{GNCCP}, denoted by GNCCP\_PGM. (added in last version)

The algorithms are implemented by Matlab R2011 on a personal computer with a $3.07$ GHz CPU (two core) and $2.00$ Gb RAM, using mex (dll) files \emph{lpsolve} toolbox\footnote{Applicable at http://sourceforge.net/projects/lpsolve/files/lpsolve/} and the \emph{rectangular assignment} toolbox\footnote{Applicable http://www.mathworks.com/matlabcentral/fileexchange/6543} for the linear programming problem (\ref{linear_programming}).

\subsection{On synthetic data}
\subsubsection{Experimental settings}
In this experiment, the accuracies of different algorithms are first compared on randomly generated synthetic graphs. First two spatial point sets $\mathcal{G} = \{g_i\}_{i=1}^M$, $\mathcal{H} = \{h_i\}_{j=1}^N$ are randomly generated by uniform sampling, i.e. $g_i,~h_j \thicksim U(0,1)^{1\times 2}$. A partial permutation matrix $\XX^{gt} \in \RR^{M\times N}$ with $L$ $'1'$s is randomly generated as the ground truth correspondence. Then $\mathcal{G}$ is constructed by permutating $\mathcal{H}$ with $\XX^{gt}$ as \ben g_i = h_j + \eta,\hspace{0.5cm} \eta\thicksim N(0,\sigma^2),\hspace{0.5cm}\text{if }\XX^{gt}_{ij} = 1, \enn where $\eta$ is the additive gaussian noise. Finally in building the graphs, distances between points are utilized as the edge weights. And the adjacency, i.e., the graph structure, is built in a sparse manner by adjusting the edge density. The graph structure is disturbed by the noise $\eta$ following a similar way in \cite{path}. Specifically, $\frac{1}{2}\sigma\#Edge$ edges are randomly added to and removed from each sparse graph, where $\#Edge$ denotes the number of edges. The adjacency matrices $\AAA_G$, $\AAA_H$, and the affinity matrix are then obtained, where the affinity matrix, required by SM, GA and PGM, is built in the same way as in \cite{pgm}. The parameter $\alpha$ is set to be $1$.

For the WCS matching, the following four scenarios are implemented.\vspace{0.1cm}\\
\textbf{Noise level }Set $M=30$, $N = M+5$, $L = M-5$, set edge density as $0.5$, and increase $\sigma$ from $0$ to $0.1$ by a step $0.01$.\\
\textbf{Problem size }Set $\sigma = 0.05$, $N = M+5$, $L = M-5$, set edge density as $0.5$, and increase $M$ from $20$ to $40$ by a step $2$.\\
\textbf{Outlier number }Set $\sigma = 0.05$, $M = 30$, $N = M+5$, set edge density as $0.5$, and decrease $L$ from $30$ to $20$ by a step $1$.\\
\textbf{Edge density }Set $\sigma = 0.05$, $M=30$, $N = M+5$, $L = M-5$, and increase the density from $0.1$ to $1$ by a step $0.1$.\vspace{0.1cm}

For the PIW problem, the following similar four scenarios are implemented.\vspace{0.1cm}\\
\textbf{Noise level }Set $N = 50$, $L = M = N-5$, set edge density as $0.5$, and increasing $\sigma$ from $0$ to $0.1$ by a step $0.01$.\\
\textbf{Problem size }Set $\sigma = 0.05$, $L = M = N-5$, set edge density as $0.5$, and increase $N$ from $40$ to $60$ by a step $2$.\\
\textbf{Outlier number }Set $\sigma = 0.05$, $N = 50$, $L = M$, set edge density as $0.5$, and decrease $L$ from $50$ to $40$ by a step $1$.\\
\textbf{Edge density }Set $\sigma = 0.05$, $N = 50$, $L = M = N-5$, and increase the density from $0.1$ to $1$ by a step $0.1$.
\vspace{0.1cm}

\subsubsection{Results}
The WCS matching performance is depicted in Fig. \ref{S_result}, from which we can draw the following six observations. \textbf{First}, generally the accuracies decrease as the noise level, problem size or outlier number increase. It is reasonable that the performances get worse as noise level and outlier number increase. A larger problem size leading to a worse performance is mainly because that the local minimum point number $\sharp min = \text{C}^L_M\text{C}^L_NL!$ for the concave relaxation increases rapidly as the vertex number increases. This makes the matching more difficult. \textbf{Second}, the algorithms achieve their highest accuracies when the edge density is about $0.5$, between $0.1$ and $1$. When the edge density is $0$, the matching problem degenerates to a pure appearance matching without structural cues. As the edge density increases, the incorporation of more structural information results in a better performance. As the edge density further increases, the accuracies by contrast decrease, which may be because excess edges associated to every vertex make the graphs less distinctive. \textbf{Third}, RLX1 and RLX2 achieve better or comparable performances with the state-of-the-art algorithm PGM. \textbf{Fourth}, RLX1 achieves a better performance than RLX2. The main reason may be $H_1(\XX)$ provides a better approximation for $H_0(\XX)$ in (\ref{obj0}). \textbf{Fifth}, the fast method introduced in Section \ref{fast_method} is effective. For instance, RLX1F achieves a comparable performance with PGM. \textbf{Sixth}, RLX1 and RLX2 outperform RLX1F and RLX2F respectively, because RLX1F and RLX2F adopt an approximate two-step scheme in each FW iteration. In spite of some accuracy loss, they have the advantage of computational efficiency, as shown below.

The time costs of different algorithms in WCS matching are compared in Fig. \ref{T_result}. It can be observed that generally the time costs with respect to the varying noise level and outlier number are relatively stable, but they are positively correlated with problem size and edge density. The time cost with respect to problem size is plotted in the \emph{log} manner in the second subfigure, which witnesses that the curve rates are respectively $5.4\pm 0.5$ for RLX1 and RLX2, $4.3\pm 0.5$ for SM, GA and PGM, and $3.5\pm 0.5$ for RLX1F and RLX2F, where the rate indicates the computational complexity, .

\befd
\centerline{\includegraphics*[width=0.45\linewidth, bb = 65 494 538 512]{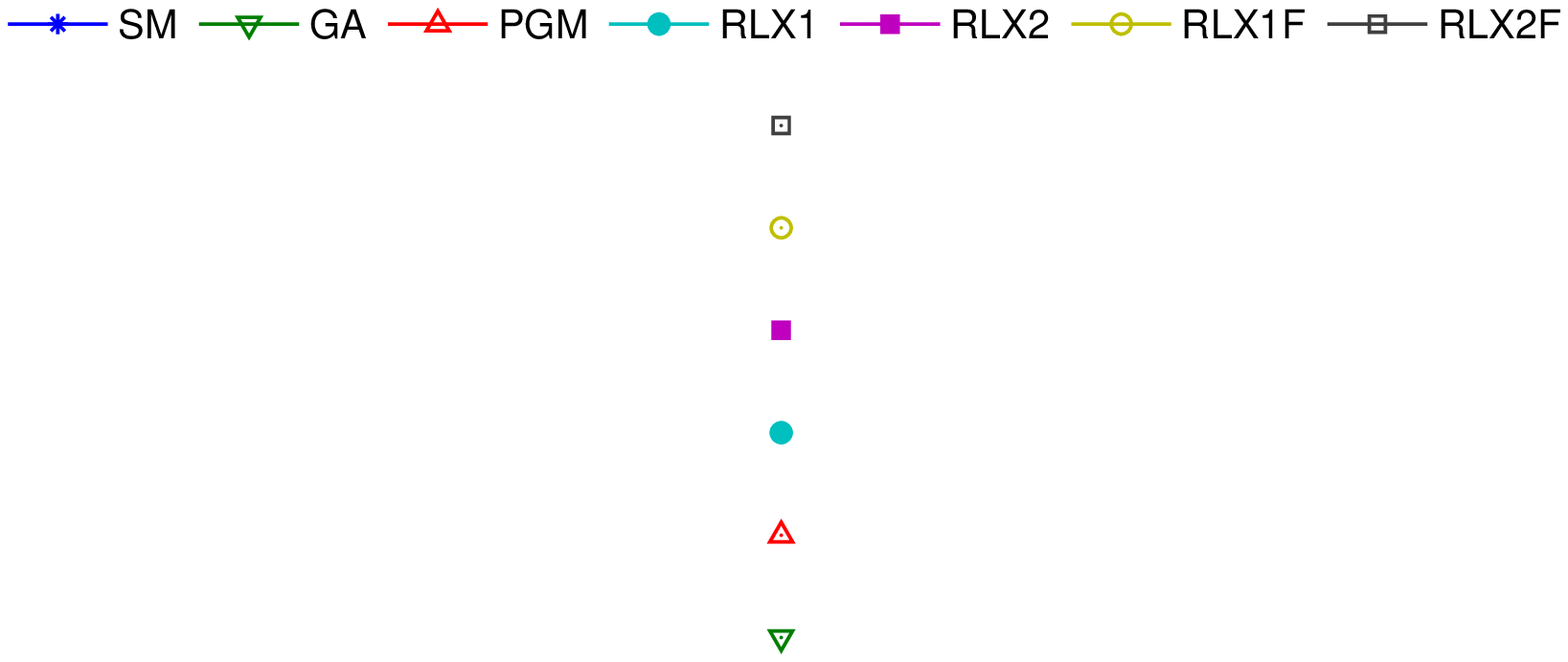}}
\centering
\includegraphics*[width=0.24\linewidth]{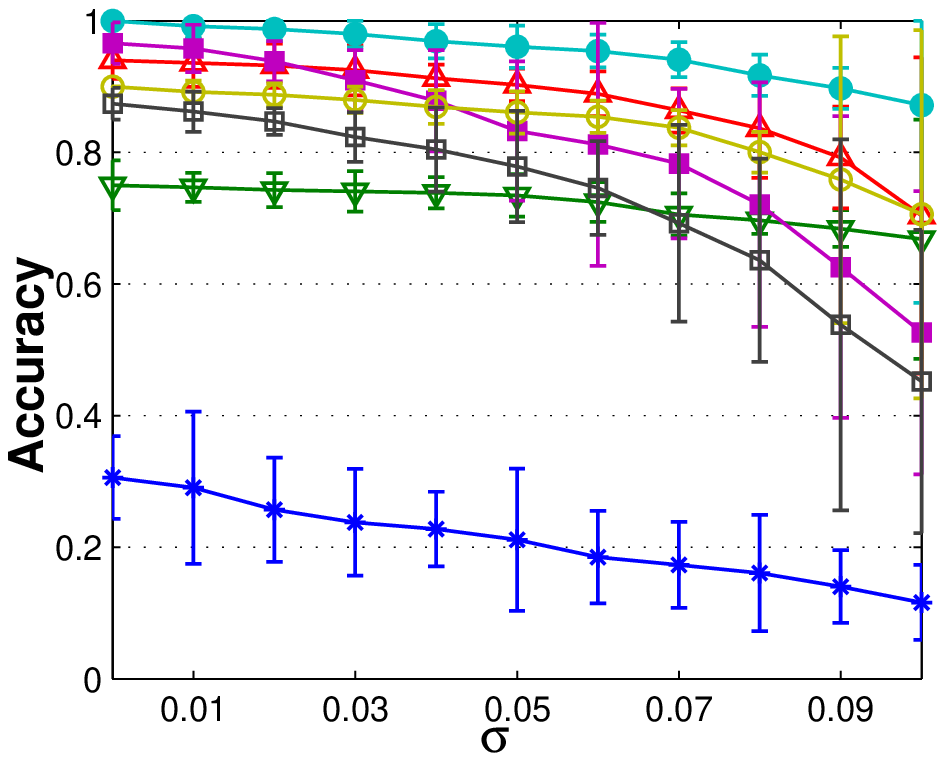}
\includegraphics*[width=0.24\linewidth]{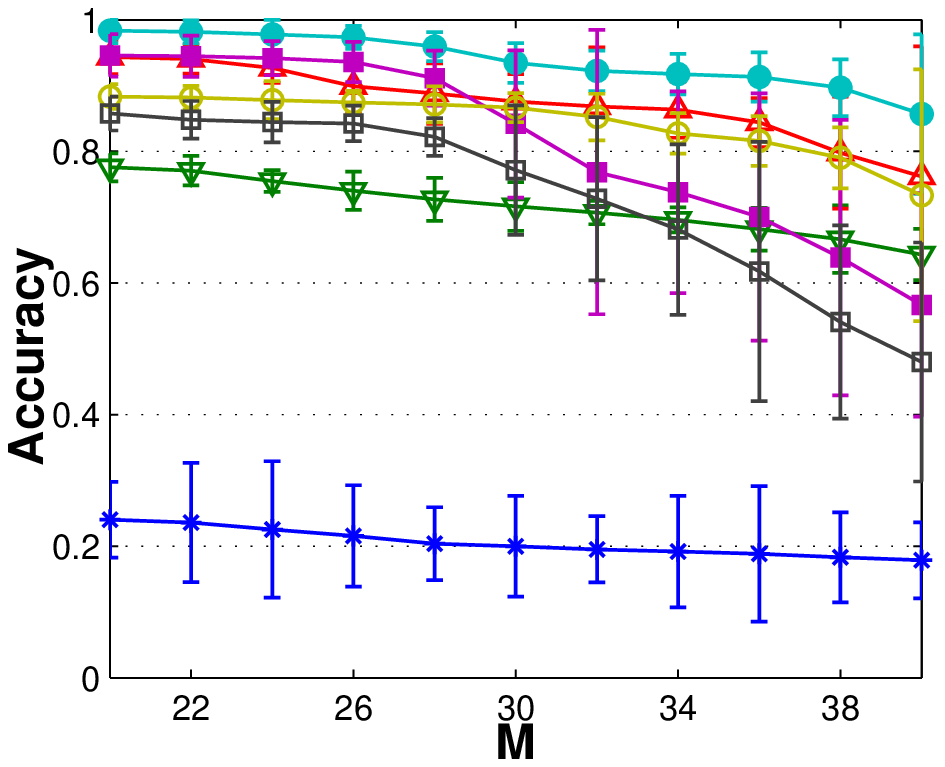}
\includegraphics*[width=0.24\linewidth]{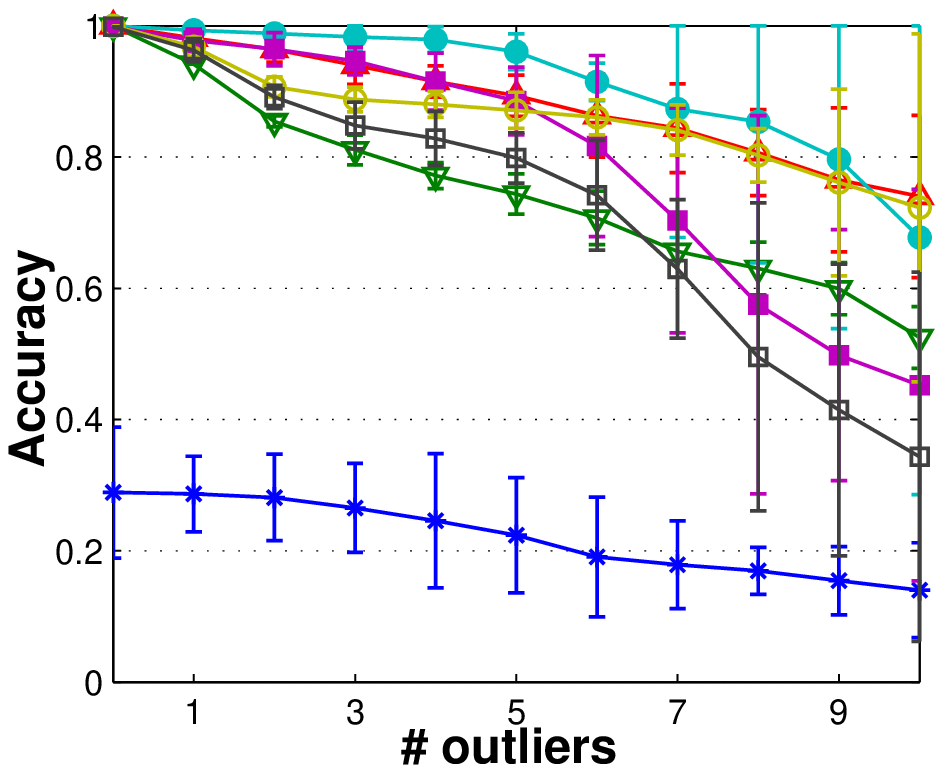}
\includegraphics*[width=0.24\linewidth]{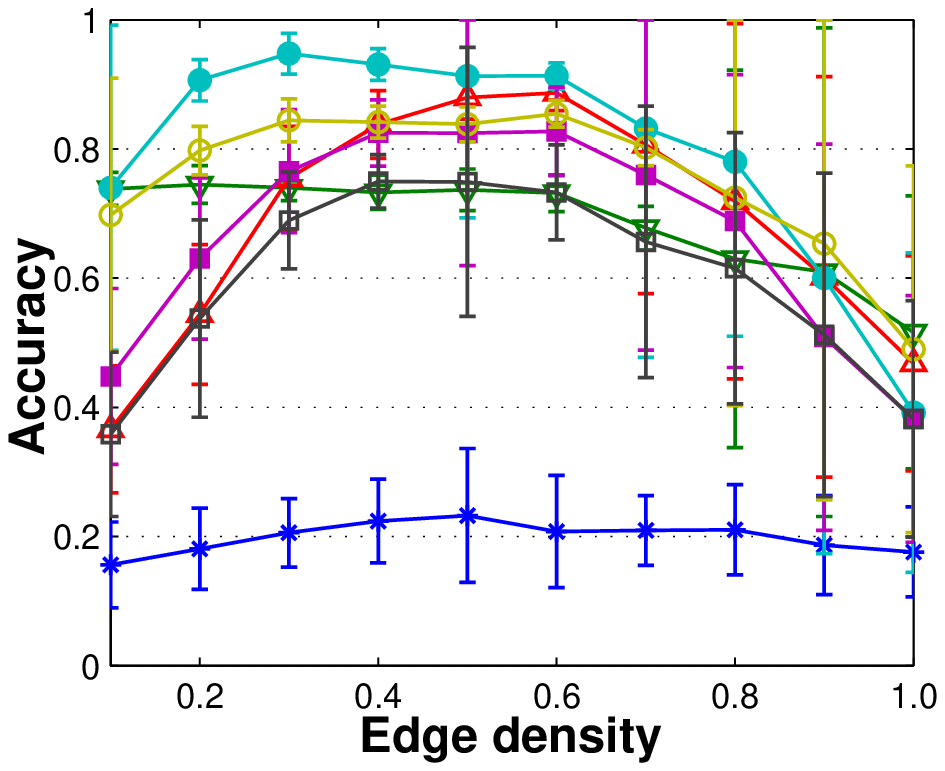}
\centering
\centerline{\includegraphics*[width=0.3\linewidth, bb = 193 494 482 511]{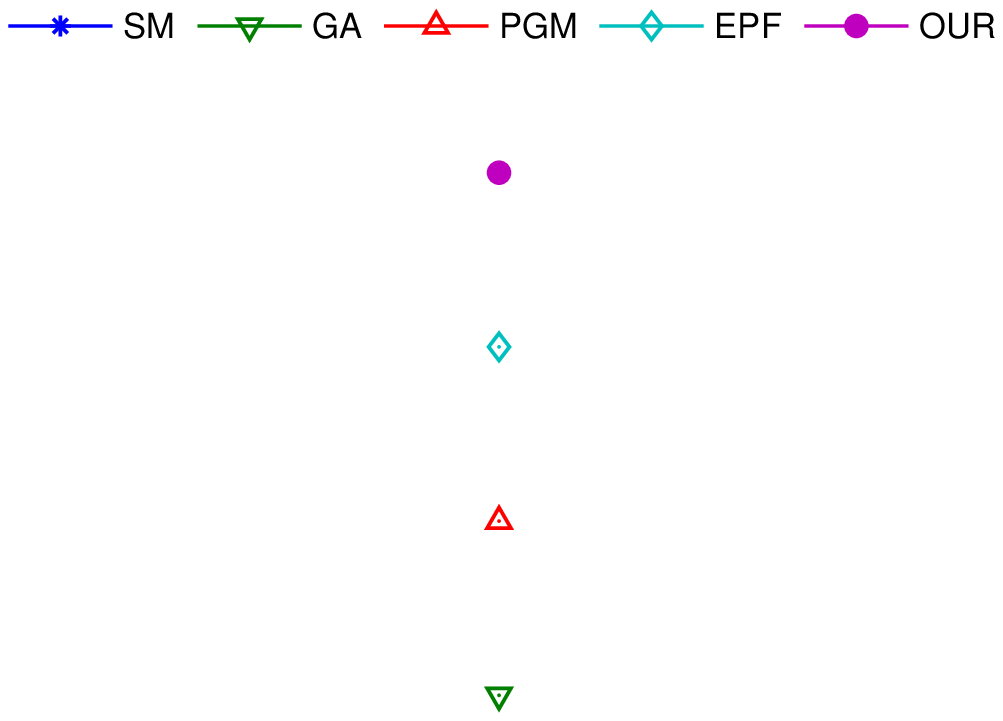}}
\includegraphics*[width=0.24\linewidth]{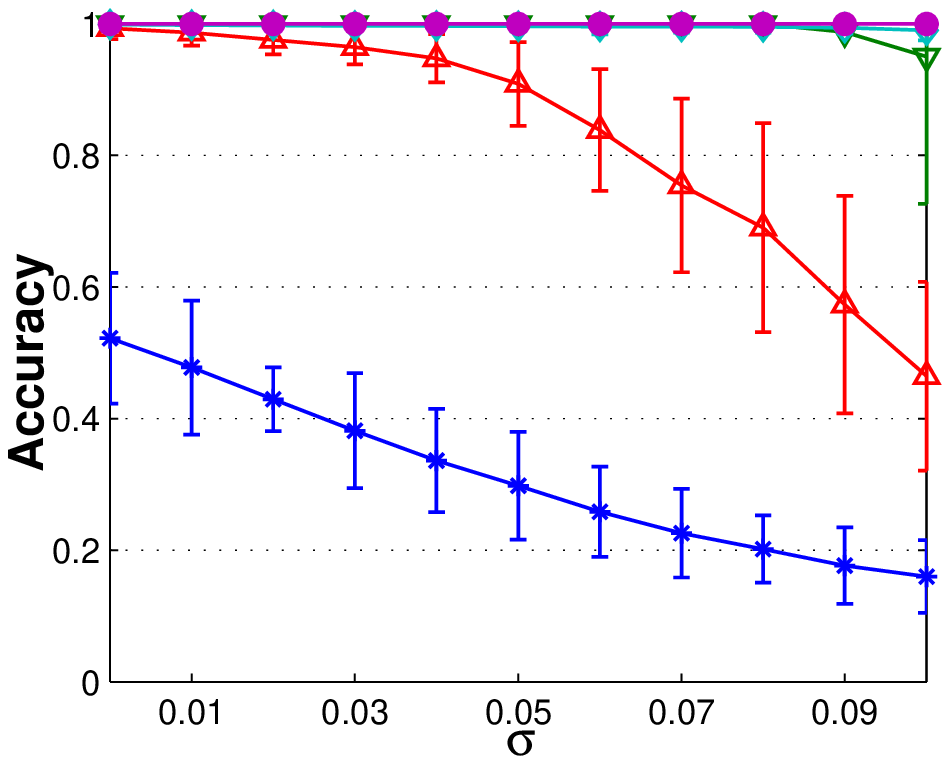}
\includegraphics*[width=0.24\linewidth]{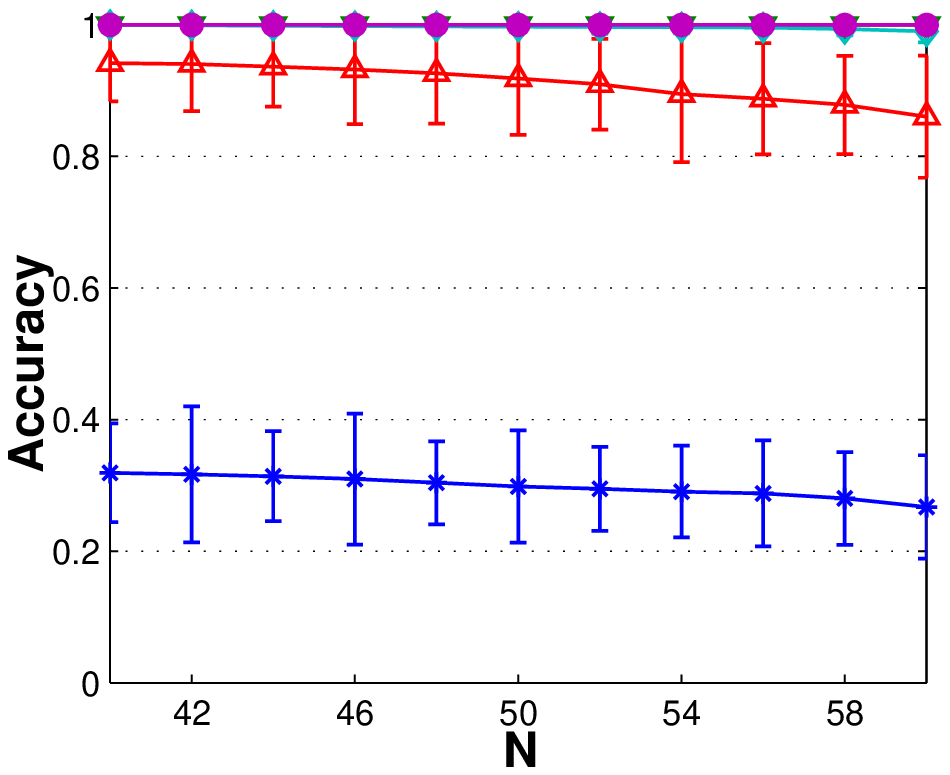}
\includegraphics*[width=0.24\linewidth]{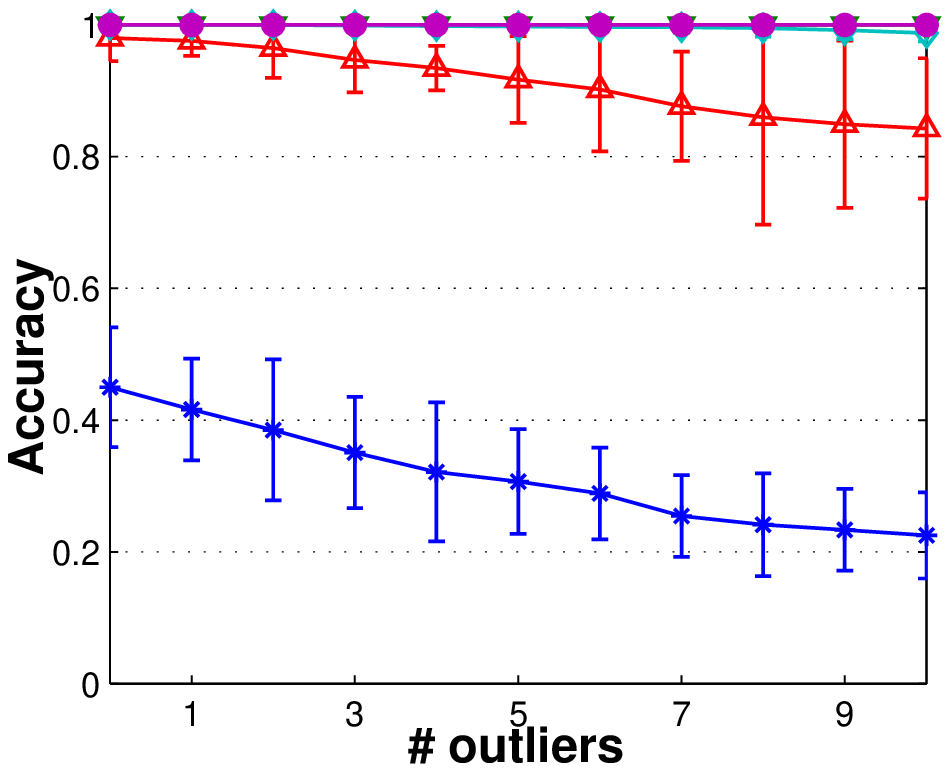}
\includegraphics*[width=0.24\linewidth]{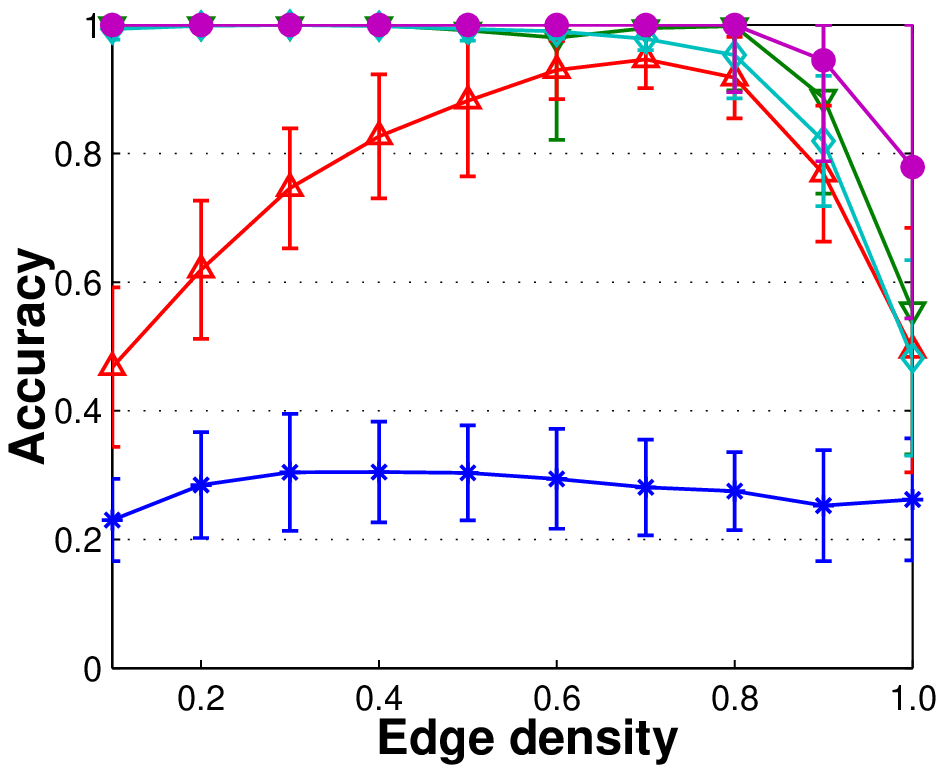}
\caption{Results on synthetic data with respect to \emph{noise level}, \emph{problem size}, \emph{outlier number} and \emph{edge density}, summarized from $30$ random runs for each fixed configuration. The WCS and PIW results are in the upper and bottom rows respectively.}\label{S_result}
\vspace{-0.5cm}
\enfd

\befd
\centerline{\includegraphics*[width=0.45\linewidth, bb = 65 494 538 512]{title1}}
\includegraphics*[width=0.24\linewidth]{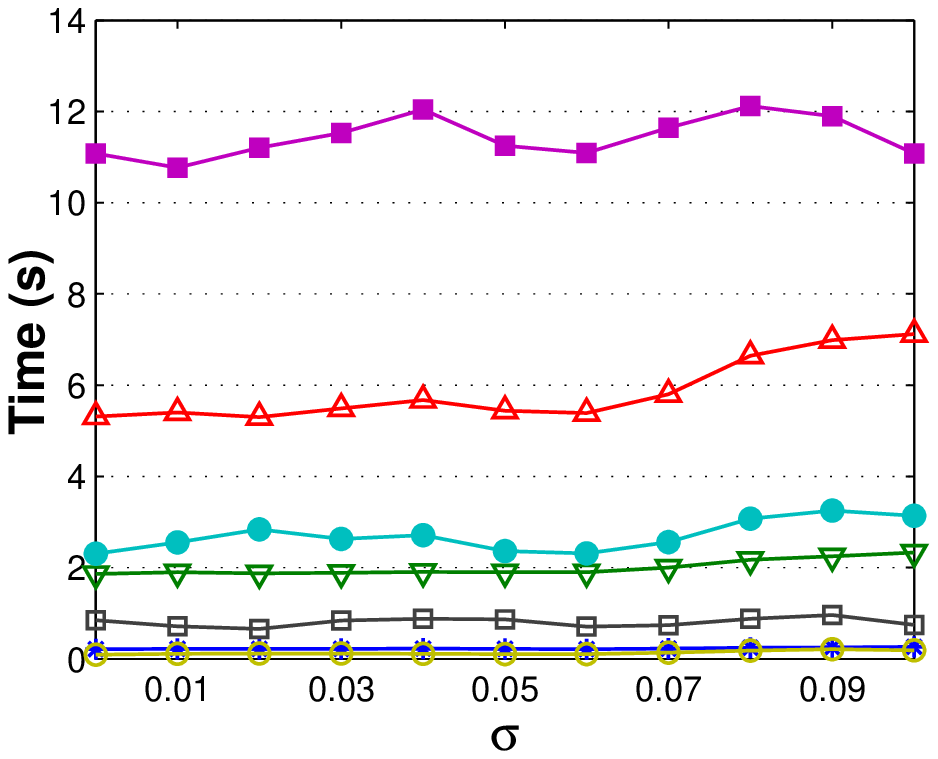}
\includegraphics*[width=0.24\linewidth]{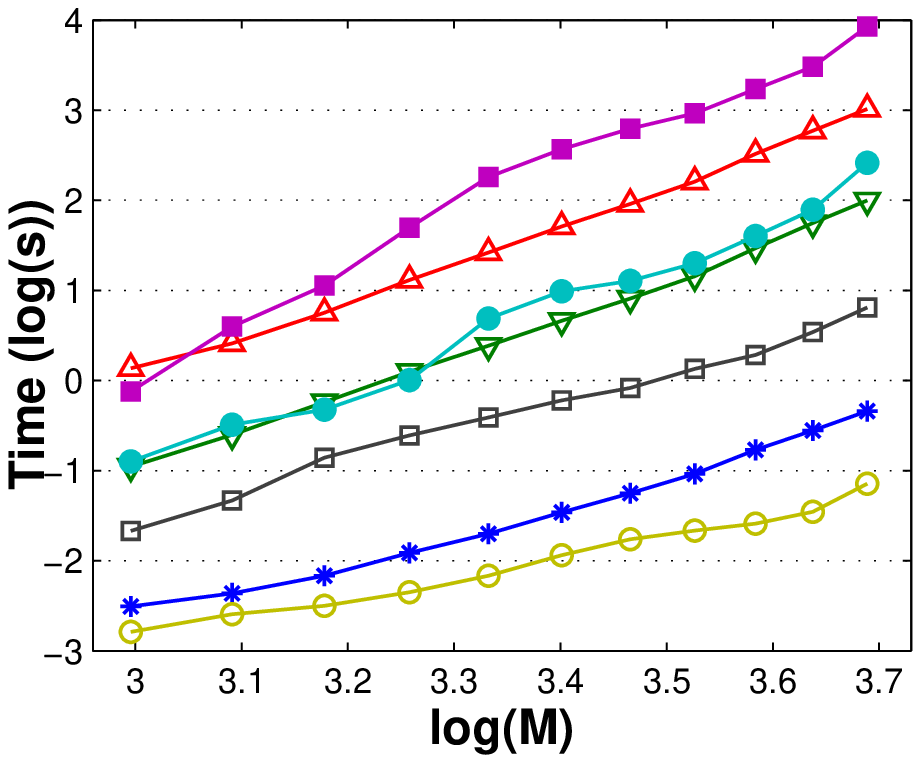}
\includegraphics*[width=0.24\linewidth]{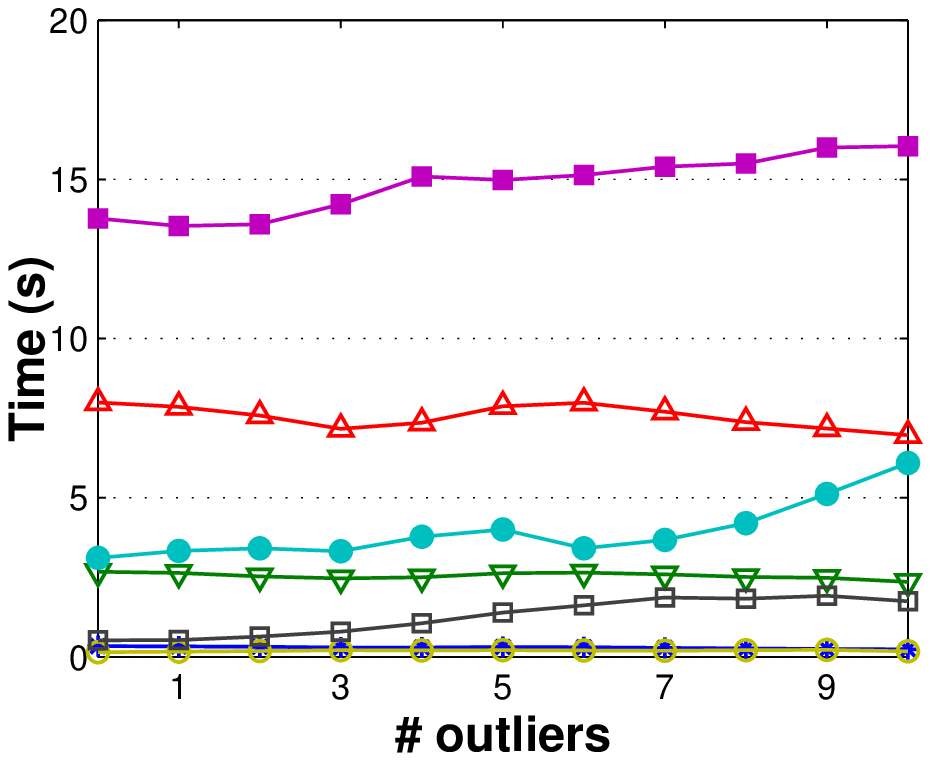}
\includegraphics*[width=0.24\linewidth]{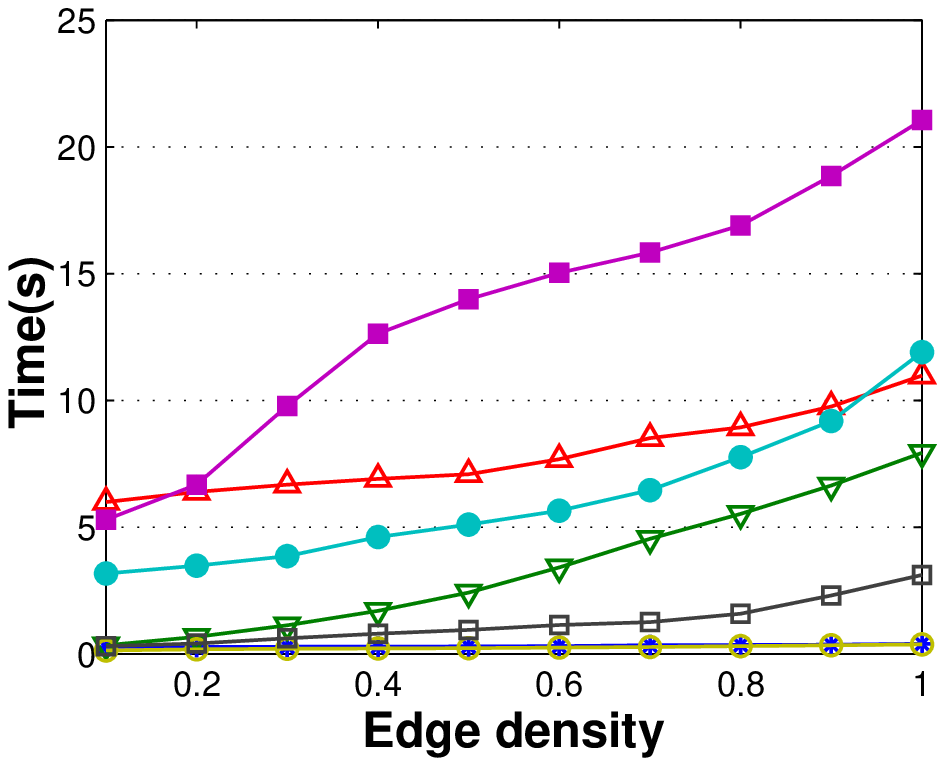}
\caption{Running time comparison in WCS matching with respect to \emph{noise level}, \emph{problem size}, \emph{outlier number} and \emph{edge density}, summarized from $50$ random runs for each fixed configuration.}\label{T_result}
\vspace{-0.6cm}
\enfd

The PIW results are also given in Fig. \ref{S_result}. All the algorithms achieve better performances on the PIW problem than on the WCS matching, implying that WCS matching is a more difficult problem. EPF transforms the subgraph matching problem into the equal-sized adjacency matrix matching by adding dummy nodes, which may however change the original problem \cite{GNCCP}. By contrast, OUR directly optimizes the subgraph matching objective, and thus achieves better results.

\subsection{On real world images}
We also apply the proposed method on a dataset fetched from Caltech256 \cite{cal256}. The dataset consists of 10 pairs of \emph{Motorbike} images and 10 pairs of \emph{Pisa} images. Each image pair is manually labeled with $60$ ground truth correspondence points. The graph structure is constructed by Delaunay triangulation. SIFT descriptor is utilized as the vertex label with $\alpha = 0.5$. The comparisons with respect to problem size and outlier number are carried out. The experimental settings are the same on the previous synthetic graph matching. The smaller number of ground truth points are randomly selected from the original ones. For instance, when the problem size is $35$ and the outlier number is $5$, only $30$ ground truth correspondence points are randomly selected from the $60$ ground truth ones. For each image pair, the outliers are randomly sampled for $10$ times. Thus for each fixed configuration, the matching are repeated for $200$ times.

The real image matching results are depicted in Fig. \ref{P_result}, which reveals that the proposed algorithms achieve better or at least comparable performances with the state-of-the-art algorithms on both WCS and PIW problems.

\bef
\centerline{\includegraphics*[width=0.9\linewidth, bb = 65 494 538 512]{title1}}
\includegraphics*[width=0.47\linewidth]{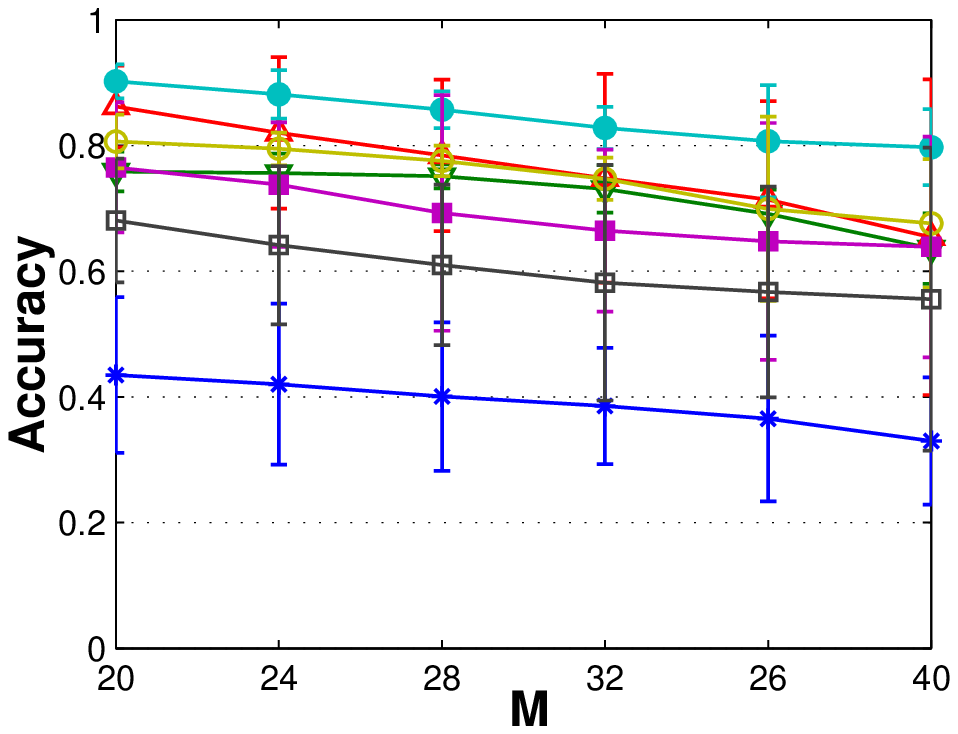}
\includegraphics*[width=0.47\linewidth]{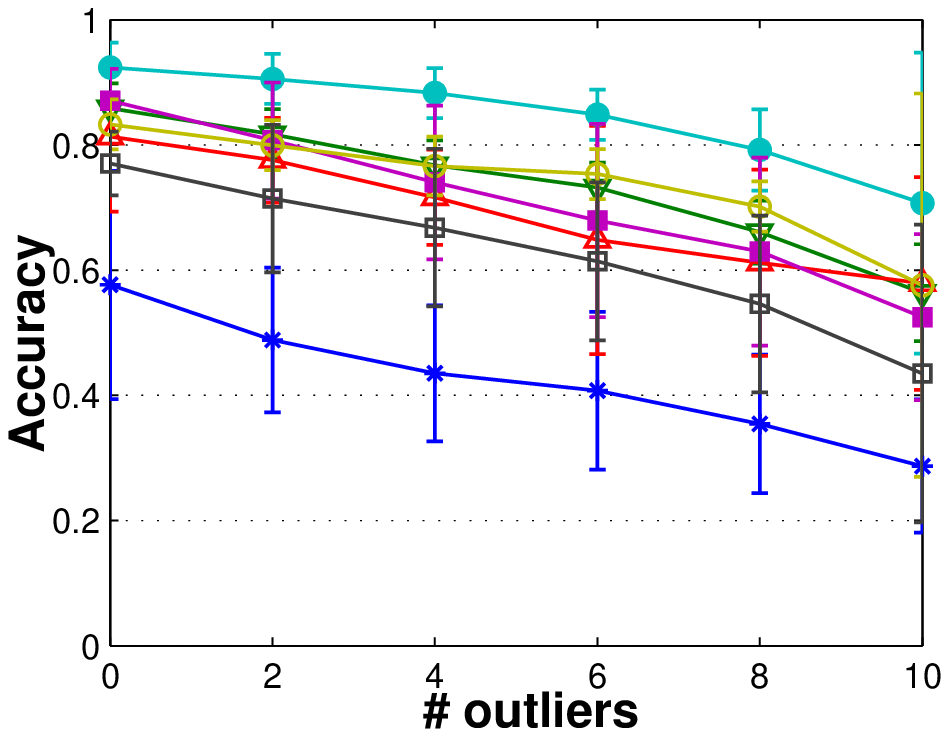}
\centerline{\includegraphics*[width=0.6\linewidth, bb = 193 494 482 511]{title2}}
\includegraphics*[width=0.47\linewidth]{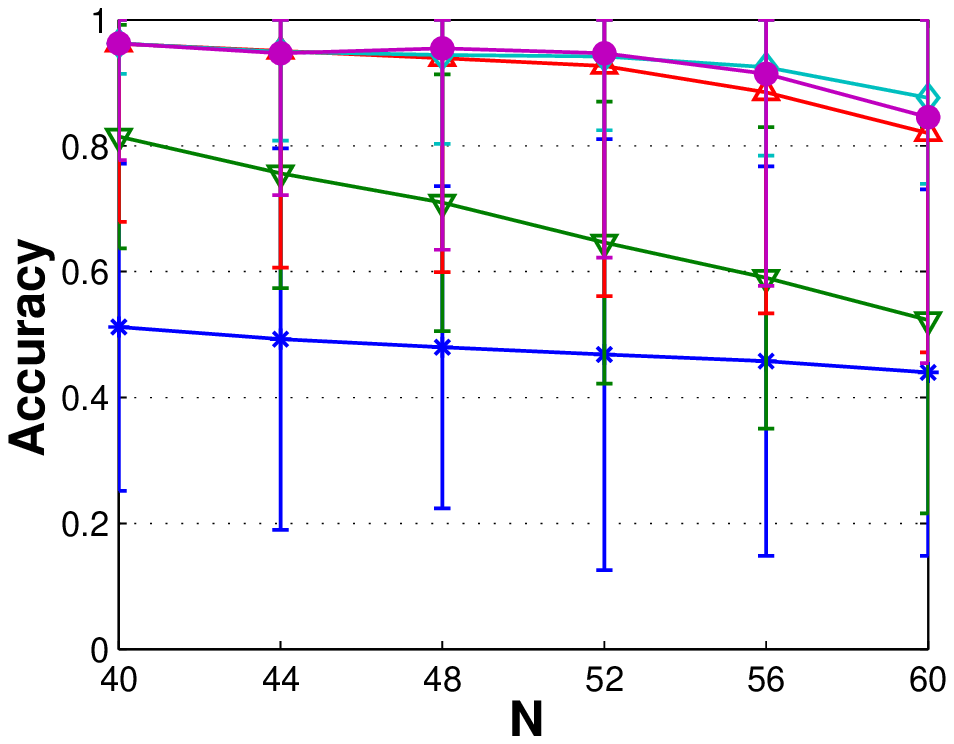}
\includegraphics*[width=0.47\linewidth]{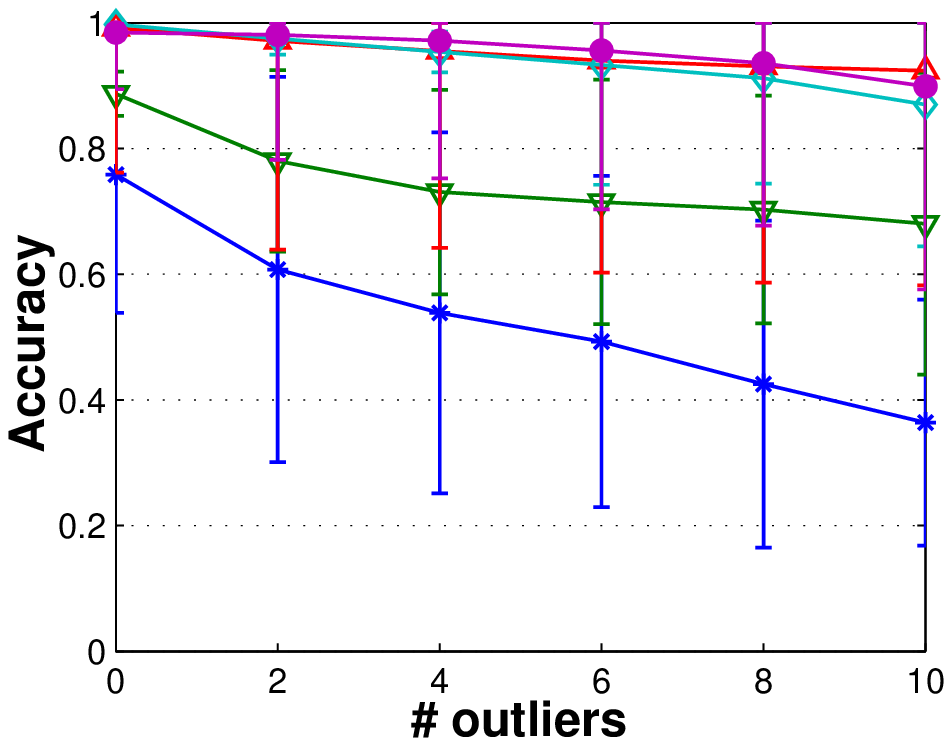}
\caption{Results on real world images with respect to \emph{problem size} and \emph{outlier number}, summarized from $200$ random runs for each fixed configuration. The WCS and PIW results are in the upper and bottom rows respectively.}\label{P_result}
\vspace{-0.5cm}
\enf

\section{Conclusion and Future Works}\label{sec:conclusion}
A novel weighted common subgraph matching algorithm has been proposed in this paper. Different from the commonly used two-step strategy, the proposed WCS matching algorithm can directly find out the most similar subgraphs of a specified size within two labeled weighted graphs. A limitation of the proposed algorithm is that the common subgraph size must be pre-specified. Though the specification may be convenient in some applications \cite{robust_point}, sometimes we may prefer to an automatic selection, which is one of our future works. On the other hand, the proposed method can find only one optimal solution, i.e. one pair of the most similar subgraphs. A general formulation and approach for solving the multiple solution problem is another future work.
\bibliographystyle{IEEEbib}
\bibliography{IEEEabrv,ref}
\end{document}